\documentclass[conference]{IEEEtran}

\usepackage{graphicx,times,psfig,amsmath} 

\begin{document}

\title{Self-control Dynamics for Sparsely Coded Networks with Synaptic Noise}

\author{\authorblockN{D. Boll\'e and R. Heylen}
\authorblockA{Institute for Theoretical Physics\\
Katholieke Universiteit Leuven\\
B-3001, Leuven, Belgium\\
E-mail: desire.bolle@fys.kuleuven.ac.be}}

\maketitle

\begin{abstract}
For the retrieval dynamics of sparsely coded attractor associative memory 
models with synaptic noise  the inclusion of a macroscopic time-dependent 
threshold is studied. It is shown that if the threshold is chosen 
appropriately as a function of the cross-talk noise and  of the activity 
of the memorized patterns, adapting itself automatically in the course of
the time evolution, an autonomous functioning of the model is guaranteed.
This self-control mechanism considerably improves the quality of the 
fixed-point retrieval dynamics, in particular the storage capacity, 
the basins of attraction and the mutual information content.
\end{abstract}

\section{Introduction}

Efficient neural network modelling requires an autonomous functioning 
independent from external constraints or control mechanisms. For 
fixed-point retrieval by an attractor associative memory model this 
requirement is mainly expressed by the robustness of its learning 
and retrieval capabilities against external noise, against
malfunctioning of some of the connections and so on. Indeed, a
model which embodies this robustness is able to perform as a
content-adressable memory having large basins of attraction for the
memorized patterns.
Intuitively, one can imagine that these basins of attraction become
smaller when the storage capacity gets larger. This might occur, e.g.,
in sparsely coded models (Okada, 1996 and references cited therein).
Therefore, the necessity of a control of the activity of the neurons has
been emphasized such that the latter stays the same as the activity of the
memorized patterns during the recall process.
This has led to several discussions imposing external constraints on the
dynamics. However, the enforcement of such a constraint at every time
step destroys part of the autonomous functioning of the network.    
To solve this problem, quite recently,  a self-control mechanism has been
introduced in the dynamics through the introduction of a time-dependent 
threshold in the transfer function (Dominguez \& Boll\'e, 1998; Boll\'e,
Dominguez \& Amari 2000). This threshold is determined as a 
function of both the cross-talk noise and the
activity of the memorized patterns in the network and adapts itself in
the course of the time evolution.

Up to now only neural network models without synaptic noise have been
considered in this context. The purpose of the present work is precisely
to generalise this self-control mechanism when synaptic noise is allowed.

\section{The model}
Let us consider a network of $N$ binary neurons. At a discrete time step 
$t$  the neurons $\sigma_{i,t}  \in \{0,1\}, \,\,\, i=1, \ldots, N$ are
updated synchronously according to the rule
\begin{equation}
     \sigma_{i,t+1}= F_{\theta_{t}, \beta}(h_{i,t}), \,\,\,
          h_{i,t}= \sum_{j(\neq i)}^{N}J_{ij}(\sigma_{j,t}-a)\, ,
      \label{2.si}
\end{equation}
where $J_{ij}$ are the synaptic couplings, $a$ is the activity of the
memorized patterns and $h_{i,t}$ is usually called the ``local field'' 
of neuron $i$ at time $t$. 
In general, the transfer function $F_{\theta_{t}, \beta}$ can be a monotonic
function with $\theta_{t}$ a time-dependent threshold. Later on it will
be chosen as
\begin{equation}
F_{\theta_{t},\beta}(x)=\frac{1}{2}[1+ \tanh(\beta(x-\theta_t))]\, .
   \label{transfer}
\end{equation}
The ``temperature''  $\beta=1/T$ controls the thermal fluctuations,
which  are a measure for the synaptic noise
(Hertz et al., 1991). In the sequel, for theoretical simplicity in the
methods used, the number of neurons $N$ will be taken to be sufficiently
large.

The synaptic couplings $J_{ij}$ themselves are determined
by the covariance rule
\begin{equation}
    J_{ij}= \frac{C_{ij}}{C{\tilde a}}
           \sum_{\mu=1}^{p} (\xi^{\mu}_{i}-a)(\xi^{\mu}_{j}-a),
           \quad   {\tilde a}\equiv a(1-a)\,.
    \label{2.Ji}
\end{equation}
The memorized patterns $\xi^{\mu}_{i} \in\{0,1\}, \,\,\, \mu=1, \ldots, p$
are independent identically
distributed random variables (iidrv) with respect to $i$ and $\mu$
chosen according to the probability distribution
\begin{equation}
   p(\xi^{\mu}_{i})= a\delta(\xi^{\mu}_{i}-1)
                                 +(1-a)\delta(\xi^{\mu}_{i}).
        \label{2.px}
\end{equation} 
The coefficients $C_{ij}\in\{0,1\}$ are iidrv with probability
\begin{eqnarray}
Pr\{C_{ij}=d\}=[1-({C}/{N})] \delta_{d,0} + 
         ({C}/{N}) \delta_{d,1}   \nonumber \\
Pr\{C_{ij}=C_{ji}\}=(C/N)^2,\quad ({C}/{N})\ll 1, \quad C>0. 
\end{eqnarray}
This introduces the so-called extremely diluted
asymmetric architecture with $C$ measuring the average connectivity of the
network (Derrida et al., 1987). 

At this point we remark that the couplings (\ref{2.Ji}) are of
infinite range (each neuron interacts with infinitely many others) such
that our model allows a so-called mean-field theory approximation. This 
essentially means that we focus on the dynamics of a single neuron
while replacing all the other neurons by an average background local
field. In 
other words, no fluctuations of the other neurons are taken into
account, not even in response to changing the state of the chosen neuron.
In our case this approximation becomes exact because, crudely speaking, 
$h_{i,t}$ is the sum of very many terms and a central limit theorem can
be applied (Hertz et al., 1991). 

It is standard knowledge by now that synchronous mean-field theory
dynamics can be solved exactly for these diluted architectures 
(e.g., Boll\'e, 2004). Hence, the big advantage is that this 
will allow us to determine the precise effects from self-control in an 
exact way. We recall that the relevant parameters describing the solution
of this dynamics are  the retrieval overlap, $m^{\mu}_t$, between the
memorized pattern, $\xi^{\mu}_i$, and the microscopic network state,
$\sigma_{i,t}$, and the neural activity, $q_t$, given by,
respectively
\begin{equation}
      m^{\mu}_{t}\equiv \frac{1}{Na}
               \sum_{i}\xi^{\mu}_{i}\sigma_{i,t}\, ,
       \quad 
       q_{t}\equiv \frac{1}{N}\sum_{i}\sigma_{i,t}\, .
   \label{parmq}
\end{equation}      
We remark that the $m^{\mu}_t$ are normalized parameters within the interval
$[\,0,1]$ which attain the maximal value $1$ whenever the model succeeds
in a perfect recall, i.e., $\sigma_{i,t}= \xi^{\mu}_{i}$ for all $i$.

In order to measure the retrieval quality of the recall process, we use
the mutual information function (Boll\'e, Dominguez \& Amari, 2000;
Nadal, Brunel \& Parga, 1998; Schultz \& Treves, 1998 and references
therein). In general, it
measures the average amount of information that can be received by
the user by observing the signal at the output of a channel (Blahut,
1990; Shannon, 1948).
For the recall process of memorized patterns that we are discussing 
here, at each time step the process can be regarded as a channel with
input $\xi_i^\mu$ and output $\sigma_{i,t}$ such that this mutual
information function can be defined as (forgetting about the pattern
index $\mu$ and the time index $t$)
\begin{eqnarray}
    &&I(\sigma_i;\xi_i)=S(\sigma_i)-
           \langle S(\sigma_i|\xi_i)\rangle_{\xi_i};
             \label{3.Is} \\
    &&S(\sigma_i) \equiv -\sum_{\sigma_i}p(\sigma_i)\ln[p(\sigma_i)],
             \label{3.Ss} \\
    &&S(\sigma_i|\xi_i)
       \equiv -\sum_{\sigma_i}p(\sigma_i|\xi_i)\ln[p(\sigma_i|\xi_i)].
             \label{3.Sx}
\end{eqnarray}
Here $S(\sigma_i)$ and $S(\sigma_i|\xi_i)$ are the entropy
and the conditional entropy of the output, respectively. 
These information entropies are peculiar to the probability 
distributions of the output.
The term $\langle S(\sigma_i|\xi_i)\rangle_{\xi_i}$ is also
called the equivocation term in the recall process.
The quantity $p(\sigma_i)$ denotes the probability distribution for the
neurons at time $t$, while $p(\sigma_i|\xi_i)$ indicates the conditional
probability that the $i-th$ neuron is in a state $\sigma_{i}$ at time $t$,
given that the $i-th$ pixel of the memorized pattern that is being retrieved
is $\xi_{i}$.
Hereby, we have assumed that the conditional probability of all the
neurons factorizes, i.e.,
 $p(\{\sigma_i\}|\{\xi_i\})=\prod_i p(\sigma_i|\xi_i)$, which is a
consequence of the mean-field theory character of our model explained
above. We remark that a similar factorization  has also been
used in Schwenker et al. (1996).

The calculation of the different terms in the expression (\ref{3.Is})
proceeds as follows. Formally writing $\langle O \rangle
\equiv \langle \langle O \rangle_{\sigma|\xi} \rangle_{\xi}=
\sum_{\xi} p(\xi) \sum_{\sigma} p(\sigma|\xi) O $ for an arbitrary
quantity $O$ the conditional probability can be obtained in a rather
straightforward way by using the complete knowledge about the system:
$\langle \xi \rangle=a, \, \langle \sigma \rangle=q, \,
\langle \sigma \xi \rangle=am, \, \langle 1 \rangle=1$.
The result reads (we forget about the index $i$)
\begin{eqnarray}
p(\sigma|\xi)&=&
     [\gamma_{0}+(m-\gamma_{0})\xi]\,\delta(\sigma-1) \nonumber\\
     &+&
     [1-\gamma_{0}-(m-\gamma_{0})\xi]\,\delta(\sigma) ,
     \nonumber\\
     \gamma_{0}&=& \frac{q-am}{1-a}
\label{3.ps}
\end{eqnarray}
One can simply verify that this satisfies the averages
\begin{eqnarray}
  m=\frac{1}{a}\langle\langle
           \sigma \xi \rangle_{\sigma|\xi}\rangle_{\xi}
      \qquad
  q=\langle\langle\sigma\rangle_{\sigma|\xi}\rangle_{\xi}
          \label{3.ms}
\end{eqnarray}
and those are precisely equal, for large $N$, to the parameters $m$ and
$q$ mentioned above (Eq.~(\ref{parmq})).  
Using the probability distribution of the patterns (Eq.(\ref{2.px})), we
furthermore obtain
\begin{equation}
    p(\sigma)\equiv\sum_{\xi}p(\xi)p(\sigma|\xi)=
     q\delta(\sigma-1)+(1-q)\delta(\sigma).
   \label{3.px}
\end{equation}
Hence the expressions for the entropies defined above become
\begin{eqnarray}
   &&S(\sigma)= -q\ln q - (1-q)\ln(1-q),\,\,
         \\
   && \langle S(\sigma|\xi)\rangle_{\xi}=
    -a[m\ln(m)+ (1-m)\ln(1-m)]
        \nonumber\\
    && \hspace*{0.7cm} 
       -(1-a)[ \gamma_0 \ln \gamma_0 + (1-\gamma_0)\ln(1-\gamma_0)].
    \label{3.Hs}
\end{eqnarray}
Recalling eq. (\ref{3.Is})  this completes the calculation of the mutual 
information content of the present model. 

\section{Self-control dynamics}
It is standard knowledge (e.g., Derrida et al., 1987; Boll\'e, 2004)
 that the synchronous dynamics for diluted architectures 
can be solved exactly following the method based upon a signal-to-noise
analysis of the local field (\ref{2.si}) (e.g., Amari, 1977; Amari \&
Maginu, 1988; Okada, 1996; Boll\'e, 2004 and references therein). 
Without loss of generality we focus on the recall of one pattern, say
$\mu=1$, meaning that only $m^1_{t}$ is macroscopic, i.e., of order $1$
and the rest of the patterns causes a cross-talk noise at each time
step of the dynamics.
Supposing that the initial state of the network model,
$\{\sigma_{i,0}\}$, is a
collection of iidrv with mean zero and neural activity $q_0$ and correlated
only with memorized pattern $1$ with an overlap $m^1_0$, then the full time 
evolution can be shown to be given by
\begin{eqnarray}
  m_{t+1}^1=
     \langle F_{\theta_{t},\beta}[(1-a)M^1_{t}+\omega_{t}] \rangle_{\omega}
     \label{3.M1}\\
  q_{t+1}=a m_{t+1}^1 +
         (1-a)\langle F_{\theta_{t},\beta}(-aM^1_{t}
                       +\omega_{t}) \rangle_{\omega}\, ,
              \label{3.Q}
\end{eqnarray}
with 
\begin{equation}
 M^1_{t}=\frac{m^1_{t}-q_{t}}{1-a}, 
\end{equation}
where we have averaged over the 
first  pattern $\xi^1$ and where the angular brackets indicate that we
still have to average over the residual (cross-talk) noise $\omega_{t}$
which can be written as 
\begin{equation}
\omega_{t}=[\alpha Q_{t}]^{1/2} {\cal N}(0,1),
  \quad Q_{t}=(1-2a)q_{t} + a^2
\end{equation}
with   ${\cal N}(0,1)$ a Gaussian random
variable with mean zero and variance unity and the (finite) loading
defined by $p= \alpha C$. Recalling the specific form
of the transfer function (\ref{transfer}) we explicitly have
\begin{eqnarray}
 &&\hspace*{-1.2cm}
 \langle F_{\theta_{t},\beta}[-aM^1_{t}+\omega_{t}] \rangle_{\omega}
     \nonumber\\
  && \hspace*{-1cm}  
  = \int_{-\infty}^{\infty} 
     \frac{dy \,\,e^{-y^2/ \alpha Q_t}}{2\sqrt{2\pi \alpha Q_t}} 
 [1+ \tanh[\beta (-aM_t - \theta_{t} + y)]]
 \label{transtemp}
 \end{eqnarray}
and an analogous expression  for $\langle 
F_{\theta_{t},\beta}[(1-a)M^1_{t}+\omega_{t}] \rangle_{\omega}$.

Of course, it is known that the quality of the recall process is
influenced by the cross-talk noise at each
time step of the dynamics. A novel idea is then to let the network
itself autonomously counter this cross-talk noise at each time step by
introducing an adaptive, hence time-dependent, threshold. This has been
studied for neural network models at zero temperature, i.e., without
synaptic noise where $F_{\theta_{t}, \beta=\infty}(x)=\Theta(x-\theta_t)$.
For sparsely coded models, meaning that the pattern activity $a$ is very
small and tends to zero for $N$ large, it has been found 
(Dominguez \& Boll\'e, 1998; Boll\'e, Dominguez \& Amari, 2000) that 
\begin{equation}
    \theta_{t}(a)=c(a)\sqrt{\alpha Q_{t}}, \quad c(a)=\sqrt{-2 \ln(a)}
     \label{2.tt}
\end{equation}
makes the second term on the r.h.s of Eq.(\ref{3.Q}) asymptotically vanish
 faster than
$a$ such that $q \sim a$. 

It turns out that the inclusion of this
self-control threshold  considerably improves the quality
of the fixed-point retrieval dynamics, in particular the storage capacity, 
the basins of attraction and the information content. As an example we
present in Fig.~1 the basin of attraction for the whole retrieval phase
$R$ for the self-control model with $\theta_{sc}$ given by
Eq. (\ref{2.tt}) and initial value $q_0=0.01=a$, compared with a model where
 the threshold $\theta_{opt}$
is selected for every loading $\alpha$ by hand in an optimal way meaning
 that the
information content $i=\alpha I$ is maximized. The latter is non-trivial
because it is even rather difficult, especially in the limit of sparse
coding, to choose  a threshold interval by hand such that $i$ is 
non-zero. The basin of attraction is clearly enlarged  with this
self-control threshold choice and even near
the border of critical storage the results are still improved. For more details
we refer to Dominguez \& Boll\'e (1998) and  Boll\'e, Dominguez \& Amari
(2000).
\begin{figure}[htp]
\centering
\includegraphics[width=7cm]{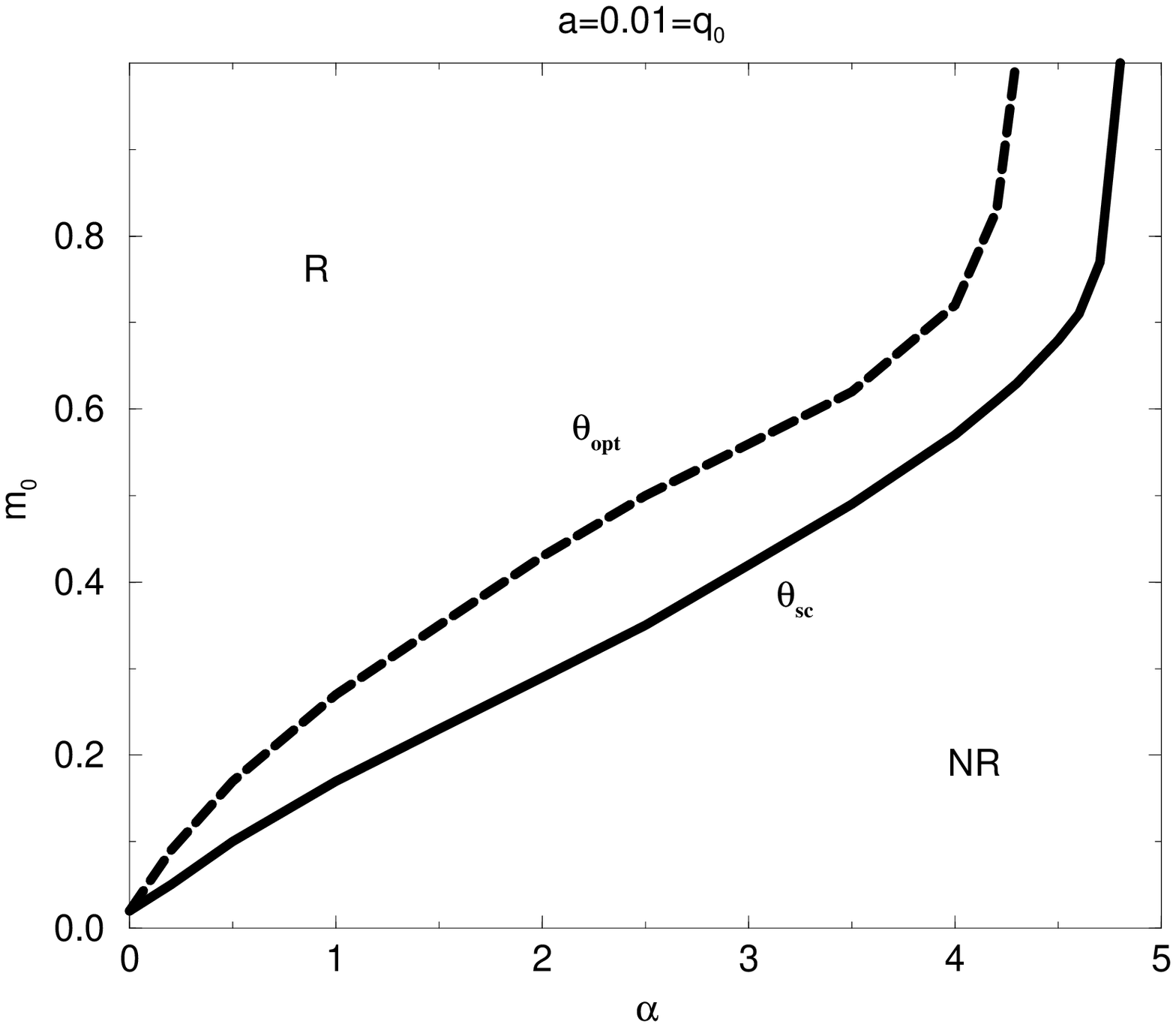}
\caption{ The basin of attraction as a function of $\alpha$ for
$a=0.01$ and initial $q_{0}=a$ for the self-control model
(full line) and the optimal threshold model (dashed line) at zero
temperature. }
\end{figure}
A similar
threshold also works for sparsely coded sequential patterns 
(Kitani \& Aoyagi, 1998) and even for non-sparse architectures as well
(Boll\'e \& Dominguez Carreta, 2000).

It is then worthwhile to examine whether such a self-control threshold can be 
found for networks with synaptic noise. No systematic study has been done 
in this case. The specific problem to be posed in analogy with the 
zero-temperature case is the following one. Can one determine a form for
the threshold $\theta_t$ in Eq.~(\ref{transtemp}) such that the integral 
vanishes asymptotically faster than $a$? 

In contrast with the zero-temperature case, where due to the simple form 
of the transfer function, this threshold could be determined analytically
(recall Eq.~(\ref{2.tt}), a  detailed study of the asymptotics of the 
integral in Eq.~(\ref{transtemp}) gives no satisfactory analytic solution. 
Therefore, we have designed a systematic numerical procedure through the
following steps: 
\begin{itemize}
\item Choose a small value for the activity $a'$.
\item Determine through numerical integration the threshold $\theta'$
such that 
\begin{equation}
\int_{-\infty}^{\infty} 
     \frac{dx \,\,e^{-x^2/ 2 \sigma^2}}{\sigma \sqrt{2\pi }} 
 \Theta (x- \theta) \leq a' \quad \mbox{for} \quad \theta > \theta' 
\end{equation}
for different values of the variance $\sigma^2={\alpha Q_t}$.
\item Determine, as a function of the temperature $T=1/\beta$, the value
for $\theta'_T$ such that  
\begin{eqnarray}
&&\hspace*{-2cm}
 \int_{-\infty}^{\infty} 
     \frac{dx \,\,e^{-y^2/ \sigma^2}}{2 \sigma \sqrt{2\pi }} 
  [1+ \tanh[\beta (x- \theta)]] \leq a' \nonumber \\ 
                && \mbox{for} \quad \theta > \theta' +\theta'_T.
\end{eqnarray}
\end{itemize}
The second step leads, as expected, precisely to a threshold having the
zero-temperature form Eq.~(\ref{2.tt}). The third step determining the
 temperature dependent part $\theta'_T$ leads
to the results shown in Fig.~2.
\begin{figure}[htp]
\centering
\includegraphics[width=6.5cm]{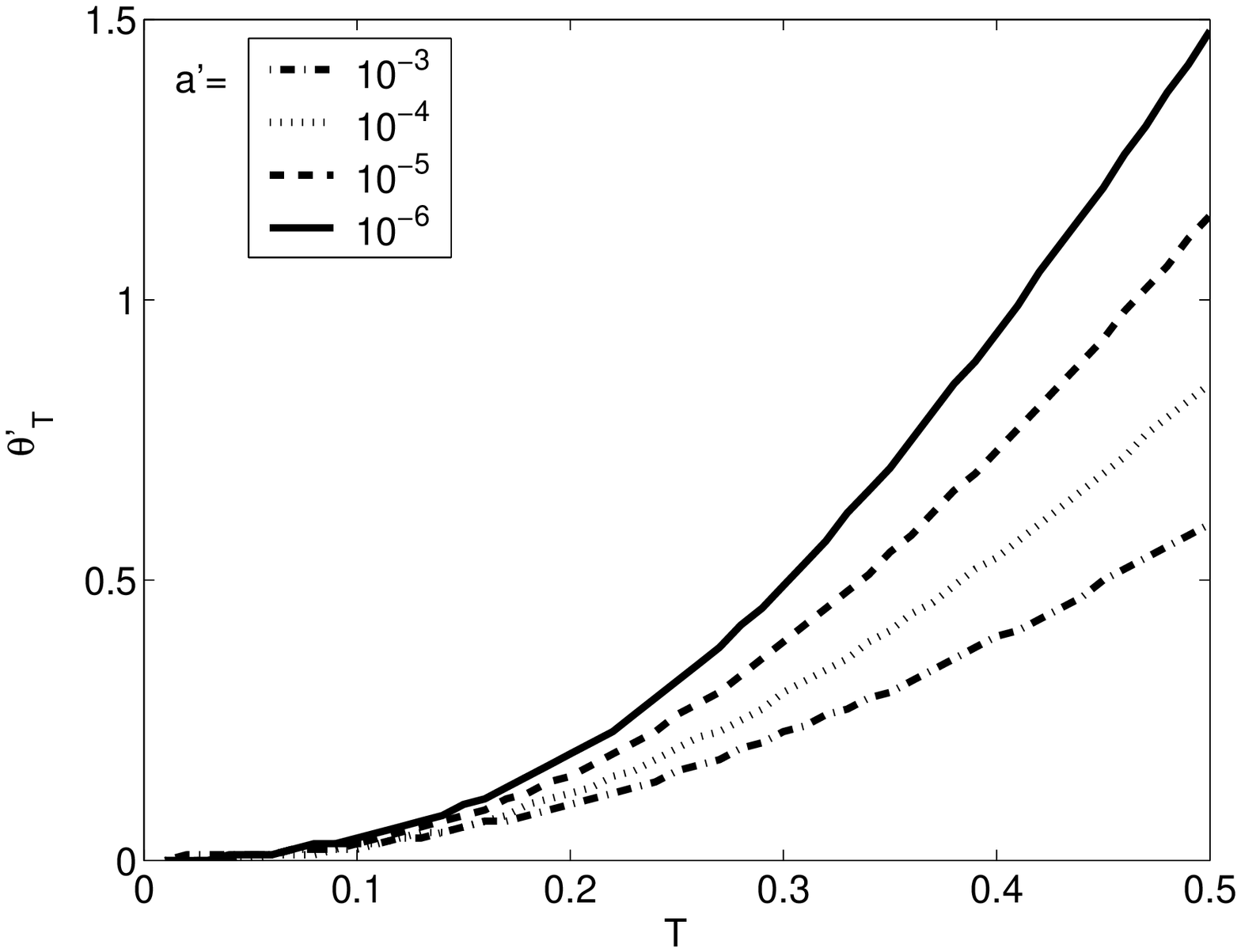}
\caption{The temperature dependent part of the threshold $\theta'_T$ as
a function of $T$ for several values of $a'$}
\label{fitten1}
\end{figure}
Intuitively it is seen that $\theta'_T$ behaves quadratically. Indeed,
making a polynomial fit of these results we find that the linear
term is negligable and that the quadratic term is of
the form $\theta'_{T}=-\frac12 \ln(a') T^2$. Furthermore, the dependence of the
coefficient of this quadratic term on the variance
is very weak. Hence, we propose the following self-control
threshold   
\begin{equation}
    \theta_{t}(a,T)=\sqrt{-2 \ln (a)\alpha Q_{t}} - \frac12 \ln(a) T^2.
     \label{threstemp}
\end{equation}
Together with Eqs.(\ref{3.M1})-(\ref{3.Q}) this relation 
describes the self-control dynamics of the network model with synaptic
noise. This dynamical threshold is again a
macroscopic parameter, thus no average must be taken over the microscopic
random variables at each time step $t$. 

At this point we want to make two remarks. 
First, for a binary layered network (Boll\'e \& Massolo, 2000) the inclusion
of a threshold of the form (\ref{2.tt}), although not designed for non-zero
temperatures, is shown to still improve the retrieval quality for low
pattern activities and low temperatures, in comparison with an optimal
threshold model analogous to the one mentioned above. Secondly, in a 
recent study of an
extremely diluted three-state neural network (Dominguez et al.,
2002) based on information theoretic and mean-field theory arguments, a
self-control threshold with a linear temperature correction term with 
coefficient $1$ has been mentioned without
any further details. In that specific model this self-control threshold 
is shown to
improve the retrieval quality for low temperatures  but it is
not specified how much of the improvement is really due to the linear
correction itself. 
\begin{figure}[htp]
\centering
\includegraphics[width=6.5cm]{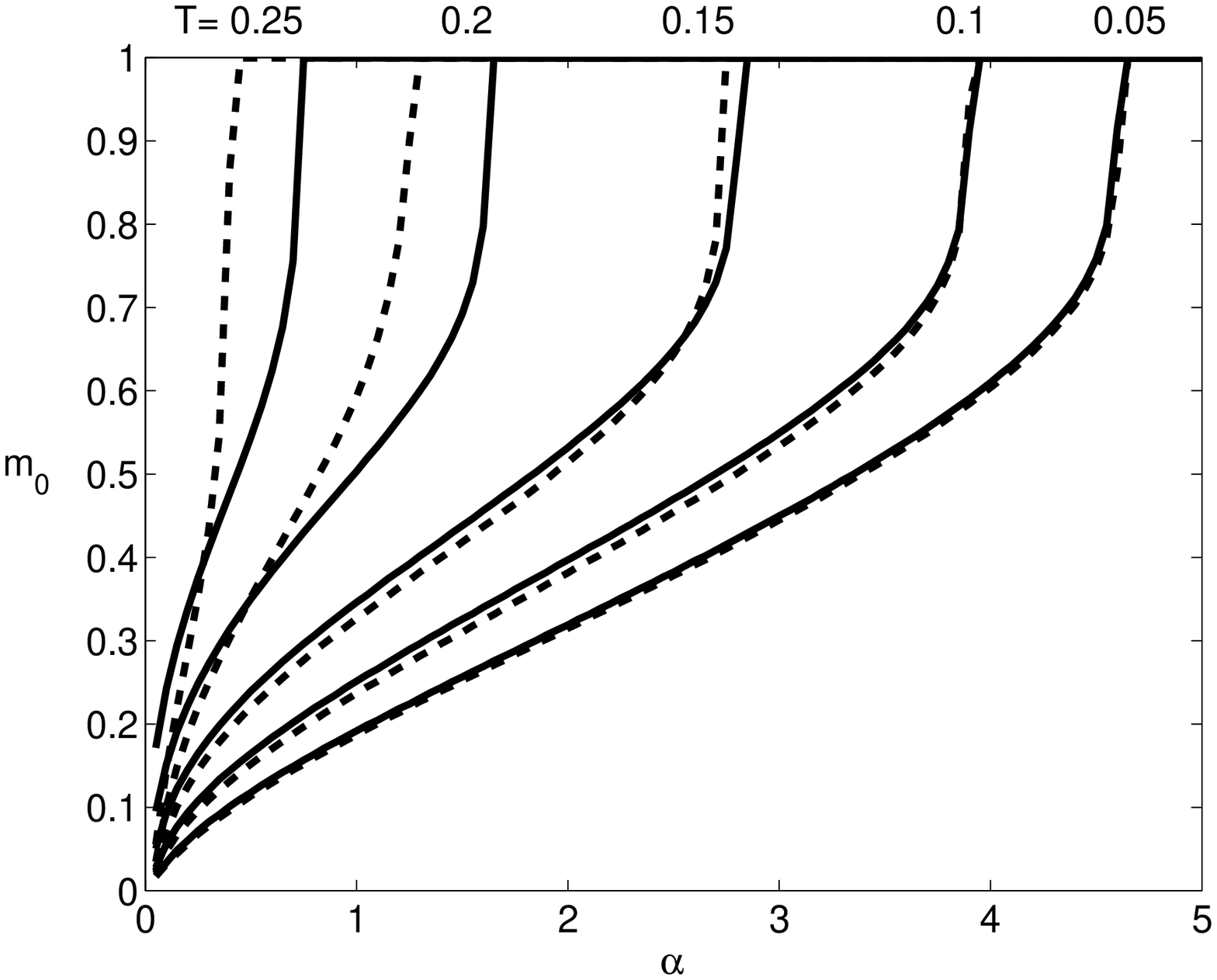}
\caption{The basin of attraction as a function of $\alpha$ for $a=0.01$
and several values of the temperature with (full
lines) and without (dashed lines) the temperature correction $\theta'_T$
in the threshold.}
\label{basins_T}
\end{figure}

We have solved this self-control dynamics,
Eqs.(\ref{3.M1})-(\ref{3.Q}) and Eq. (\ref{threstemp}), for our model with
synaptic noise, in the limit of sparse coding, numerically. In
particular, we have studied in detail the influence
of the temperature dependent part of the threshold. Of course,
we are only interested in the retrieval solutions with $M>0$ and
carrying a non-zero information~$I$. 
We remark that all numerical
calculations presented here are done for an
appropriate number of time steps (at least of the order of a few hundred) in 
order to assure that a stable equilibrium point is reached.

\begin{figure}[htp]
\centering
\includegraphics[width=6.5cm]{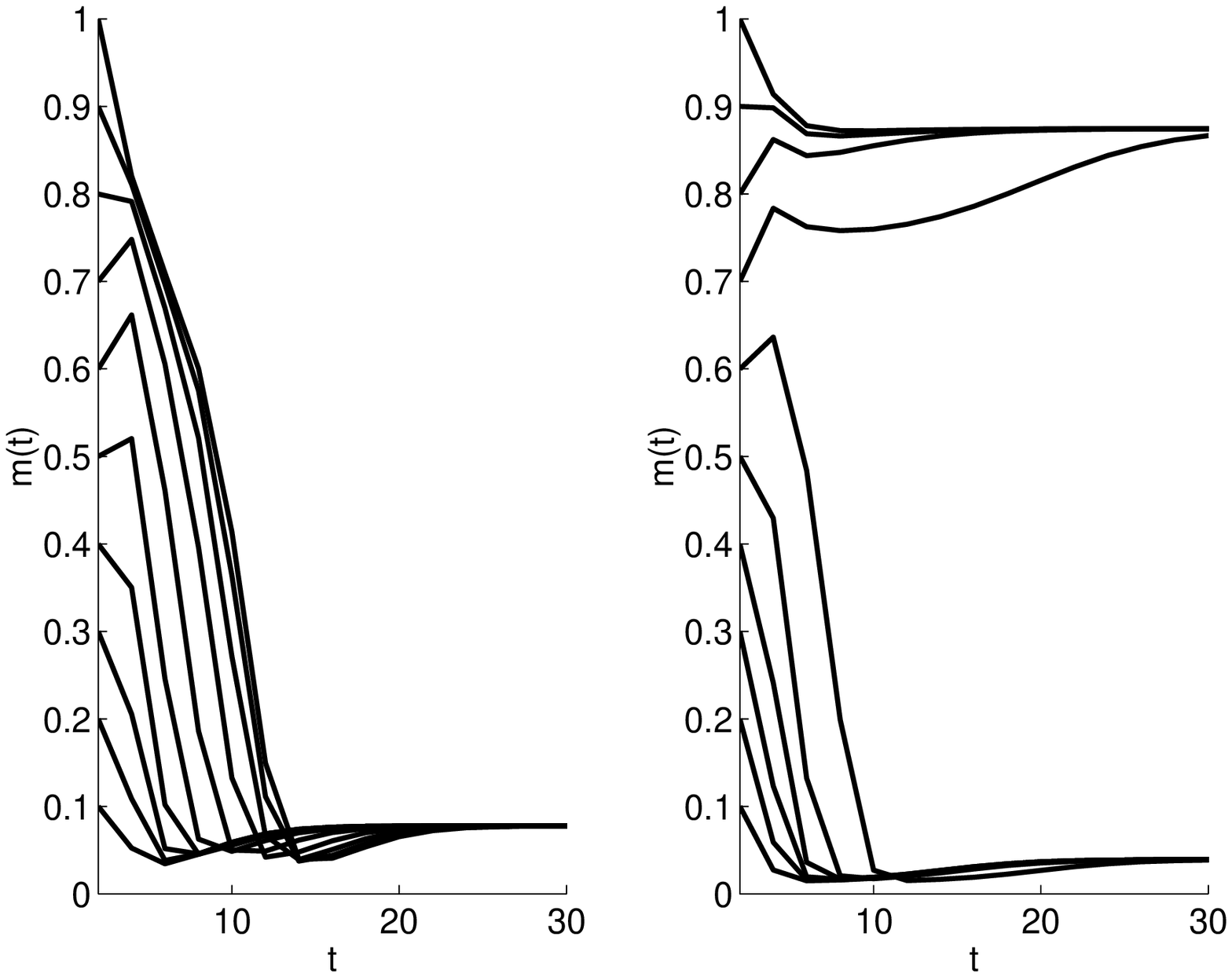}
\caption{The evolution of the overlap $m_t$ for several initial values
$m_0$ with $a=0.01$, $T=0.2$ and $\alpha=1.5$ without (left) and with
(right) the temperature correction $\theta'_T$ in the threshold.}
\label{evolutie_T}
\end{figure}

The important features of the solution are illustrated in Figs.~3-5. 
In Fig.~3 we show
the basin of attraction for the whole retrieval phase for the model with
the temperature-zero threshold (\ref{2.tt}) (dashed curves) compared to 
the model with the temperature dependent threshold (\ref{threstemp})
(full curves) (compare also Fig.~1). We see that there is no clear 
improvement for low
temperatures but there is a substantial one for higher temperatures.
Even near the border of critical storage the results are still improved
such that also the storage capacity itself is larger.

This is further illustrated in Fig.~4 where we compare the time
evolution of the retrieval overlap $m_t$ starting from several initial
values, $m_0$, for the model with (right figure) and without (left
figure) the quadratic temperature correction in the threshold. Here this 
temperature
correction is absolutely crucial to force some of the overlap
trajectories to go to the retrieval attractor $m \approx 1$. It really 
makes the
difference between retrieval and non-retrieval in the model. At this point 
we remark 
that the influence of a linear temperature correction term 
 has been examined also here but no real improvement has been
found of the results for the temperature-zero threshold. 

\begin{figure}[htp]
\centering
\includegraphics[width=6.5cm]{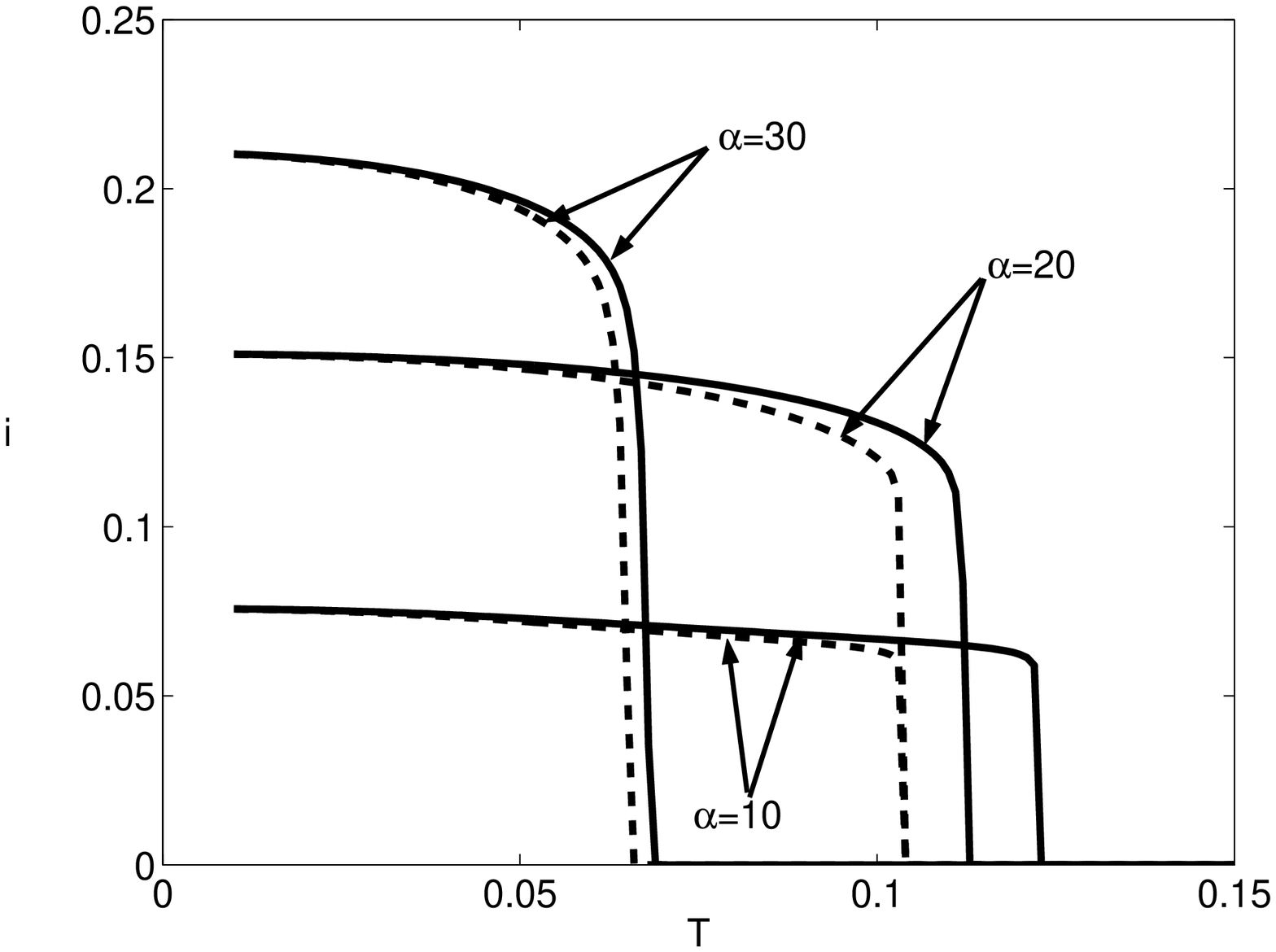}
\caption{The information content $i$ as a function of $T$ for several
values of the loading $\alpha$ and $a=0.001$ with (full
lines) and without (dashed lines) the temperature correction $\theta'_T$
in the threshold.}
\label{info_T_2}
\end{figure}

In Fig.~5 we plot the information content $i$ as a
function of the temperature for the self-control dynamics with the 
threshold (\ref{threstemp}) (full curves), respectively (\ref{2.tt})
(dashed curves). We see that,
especially for small loading $\alpha$ a substantial improvement of the
information content is obtained.

\section{Conclusions}
In this work we have generalized complete self-control in the dynamics of
sparsely coded associative memory networks to models with synaptic noise. 
We have proposed an analytic form for the relevant macroscopic threshold
consisting out of the known form for temperature zero plus a quadratic
temperature correction term dependent on the pattern activity. The 
consequences of this  self-control mechanism
on the quality of the recall process by the network have been studied.

We find that the basins of attraction of the retrieval solutions as well
as the storage capacity  are
enlarged and that the mutual information content is maximized.
This confirms the considerable improvement of the quality of recall by 
self-control, also for network models with synaptic noise. 

This allows us to conjecture  that this idea of
self-control, allowing the network to function autonomously, might be 
relevant for other architectures in the presence
of synaptic noise, and for dynamical systems in general, when
trying to improve the basins of attraction and  convergence times.

\section*{Acknowledgment}

We are indebted to S. Goossens for some contributions at the initial stages of
this work. One of the authors (DB) would like to thank D. Dominguez for 
stimulating discussions. 
This work has been supported by the Fund for Scientific Research- Flanders
(Belgium).


\begin{thebibliography}{1}
\IEEEtriggeratref{3} 


\bibitem{A77} Amari S. (1977). Neural theory and association of
concept information. {\it Biological Cybernetics}, {\bf 26}, 175-185.


\bibitem{AM88} Amari S. and Maginu K. (1988). Statistical neurodynamics
of associative memory. {\it Neural Networks}, {\bf 1}, 63-73.
    

\bibitem{B90} Blahut R.E. (1990). {\it Principles and Practice of 
  Information Theory}. Reading, MA: Addison-Wesley.

\bibitem{B04} Boll\'e D (2004).  Multi-state neural networks based upon
spin-glasses: a biased overview in {\it Advances in Condensed Matter
and Statistical Mechanics} eds. Korutcheva E and Cuerno R., Nova Science
Publishers, New-York, p. 321-349. 

\bibitem{BD00} Boll\'e D. and Dominguez Carreta D.  (2000). Mutual
information and self-control of a fully connected low-activity neural
network. {\it Physica A}, {\bf 286}, 401-416.

\bibitem{BDA00} Boll\'e D., Dominguez D.R.C. and  Amari S. (2000). Mutual
information of sparsely coded associative memory with self-control and
ternary neurons. {\it Neural Networks}, {\bf 13}, 455-462.

\bibitem{BM00} Boll\'e D. and Massolo G. (2000). Thresholds in layered neural
networks with variable activity. 
{\it Journal of Physics A}, {\bf 33}, 2597-2609.


\bibitem{DGZ87} Derrida B., Gardner E., and Zippelius A. (1987). An
exactly solvable asymmetric neural network model.
    {\it Europhysics Letters}, {\bf 4}, 167-173.

\bibitem{DB98}  Dominguez D.R.C. and Boll\'e D. (1998). Self-control in
sparsely coded networks.
     {\it Physical Review Letters}, {\bf 80}, 2961-2964.

\bibitem{DKTE02}  Dominguez D.R.C., Korutcheva E., Theumann W.K., and
Erichsen Jr. R.  (2002). Flow diagrams of the quadratic neural network.
     {\it Lecture Notes in Computer Science}, {\bf 2415}, 129-134.


\bibitem{HKP91} Hertz J., Krogh A. and Palmer R.G. (1991). {\it
Introduction to the Theory of Neural Computation}. 
 Addison-Wesley,  Redwood City.


\bibitem{KA98} Kitano K. and Aoyagi T. (1998). Retrieval dynamics of neural
networks for sparsely coded sequential patterns,
{\it Journal of Physics A}, {\bf 31}, L613-L620.

\bibitem {NBP98} Nadal J-P., Brunel N. and Parga N. (1998). Nonlinear
feedforward networks with stochastic outputs: infomax implies redundancy
reduction. {\it Network: Computation in Neural Systems}, {\bf 9}, 207-217.


\bibitem{O96} Okada M. (1996). Notions of associative memory and sparse
coding. {\it Neural Networks}, {\bf 9}, 1429-1458.



\bibitem{SSP96} Schwenker F., Sommer F.T., and Palm G. (1996). Iterative
retrieval of sparsely coded associative memory patterns. 
    {\it Neural Networks}, {\bf 9}, 445-455.

\bibitem{ST98} Schultz S. and Treves A. (1998). Stability of the
replica-symmetric solution for the information conveyed by a neural
network. {\it Physical Review E }, {\bf 57}, 3302-3310.

\bibitem{S48}  Shannon C.E. (1948). A mathematical theory for
communication. {\it  Bell Systems Technical Journal}, {\bf 27}, 379.


\end{thebibliography}
\end{document}